\documentclass{PoS}

  \usepackage{amssymb}
  \usepackage{amsmath}
    \usepackage{amsfonts}
     \usepackage{epsfig}
      \usepackage{bm}

\usepackage[section]{placeins}

\usepackage{multirow}
\usepackage{ctable}
\usepackage{booktabs}
\usepackage{array}
\usepackage{tabularx}
\usepackage{xcolor}
\usepackage{pstricks}

\newcommand{\beq}{\begin{equation}}
\newcommand{\eeq}{\end{equation}}
\newcommand{\bea}{\begin{eqnarray}}
\newcommand{\eea}{\end{eqnarray}}
\newcommand{\beqa}{\begin{eqnarray}}
\newcommand{\eeqa}{\end{eqnarray}}

\definecolor{red}{rgb}{1,0,0}

\def\be{\begin{equation}}
\def\ee{\end{equation}}
\numberwithin{equation}{section}

\title{New single- and double-parton scattering mechanisms for double charmed meson production}

\ShortTitle{Double charm meson production at the LHC}

\author{\speaker{Antoni Szczurek}\thanks{The work has been supported by the Polish National Science Center grant
DEC-2014/15/B/ST2/02528.}\\
        Institute of Nuclear
Physics, Polish Academy of Sciences, Radzikowskiego 152,\\ PL-31-342 Krak{\'o}w, Poland\\
          E-mail: \email{antoni.szczurek@ifj.edu.pl}}

\author{Rafa{\l} Maciu{\l}a\\
Institute of Nuclear
Physics, Polish Academy of Sciences, Radzikowskiego 152,\\ PL-31-342 Krak{\'o}w, Poland\\
              E-mail: \email{rafal.maciula@ifj.edu.pl}}
              
\author{Vladimir A. Saleev\\
Samara National Research University, Moscow Highway, 34, 443086, Samara, Russia\\
E-mail: \email{saleev@samsu.ru}} 

\author{Alexandra V. Shipilova\\
Samara National Research University, Moscow Highway, 34, 443086, Samara, Russia\\
E-mail: \email{alexshipilova@samsu.ru}}              

\abstract{We discuss charm meson-meson pair production recently observed by the LHCb Collaboration at $\sqrt{s}$ = 7 TeV in proton-proton scattering.
We examine double-parton scattering (DPS) mechanisms of double $c \bar
c$ production and following $cc \to D^{0}D^{0}$ hadronization as well as double $g$ and mixed
$g c\bar c $ production with $gg \to D^{0}D^{0}$ and $gc \to D^{0}D^{0}$ hadronization calculated 
with the help of the scale-dependent KKKS08 fragmentation functions.
A new single-parton scattering (SPS) mechanism of $gg$ production is also taken into consideration.
Calculated differential distributions as a function of transverse momentum $p_{T}$ of one of the $D^{0}$ mesons, pair invariant mass $M_{D^{0}D^{0}}$ and azimuthal angle $\varphi_{D^{0}D^{0}}$ distributions are confronted with the measured ones. 
The manifestation of the new SPS mechanisms with $g \to D^{0}$ fragmentation within the scale-dependent fragmentation scheme change the overall picture
suitable for standard scale-independent fragmentation where only DPS $cc \to D^{0}D^{0}$ mechanism is present.  
Some consequences of the new mechanisms are discussed.}

\FullConference{XXIV International Workshop on Deep-Inelastic Scattering and Related Subjects\\
		11-15 April, 2016\\
		DESY Hamburg, Germany}

\begin{document}
\section{Introduction}

At present, double charm production is expected to be one of the most promising channels for
studies of double-parton scattering (DPS) effects at the LHC. This was predicted \cite{Luszczak:2011zp} and further supported by the experimental observations reported by the LHCb Collaboration \cite{Aaij:2012dz}. Next, the phenomenology of $DD$ meson-meson pair production was carefully examined in the $k_t$-factorization approach and a relatively good description of the LHCb experimental data was achieved for both the total yield and the dimeson correlation observables \cite{Maciula:2013kd}. In the theoretical analyses, both, single- and double-parton scattering mechanisms were taken into consideration. The contribution of single-parton scattering (SPS) mechanism $g g \to c \bar c c \bar c$, discussed in detail in the collinear \cite{vanHameren:2014ava} and $k_t$-factorization \cite{vanHameren:2015wva} approaches, was found to be rather
small and definitely not able to describe relatively large $DD$ cross sections measured by the LHCb.

The phenomenological studies of the $DD$ pair production were based on the rather standard fragmentation scheme with scale-independent Peterson fragmentation function (FF) \cite{Peterson:1982ak}, where only $c \to D$ transition is included. However, an alternative approach is to apply     
scale-dependent FFs that undergo DGLAP evolution equations, e.g. KKKS08 model \cite{Kneesch:2007ey}, where each parton (gluon, $u,d,s,\bar u, \bar d, \bar s, c, \bar c$) can contribute to $D$ meson production. In the latter scenario, the $c \to D$ contribution is reduced by the evolution of the FF but a very important contribution from $g \to D$ fragmentation appears (see e.g. Ref.~\cite{Nefedov:2014qea}).

In this presentation we report on first investigation how important is the
gluon fragmentation mechanism for the double $D$-meson production.

\section{A sketch of the theoretical formalism}

\begin{figure}[!h]
\begin{center}
\begin{minipage}{0.24\textwidth}
 \centerline{\includegraphics[width=1.0\textwidth]{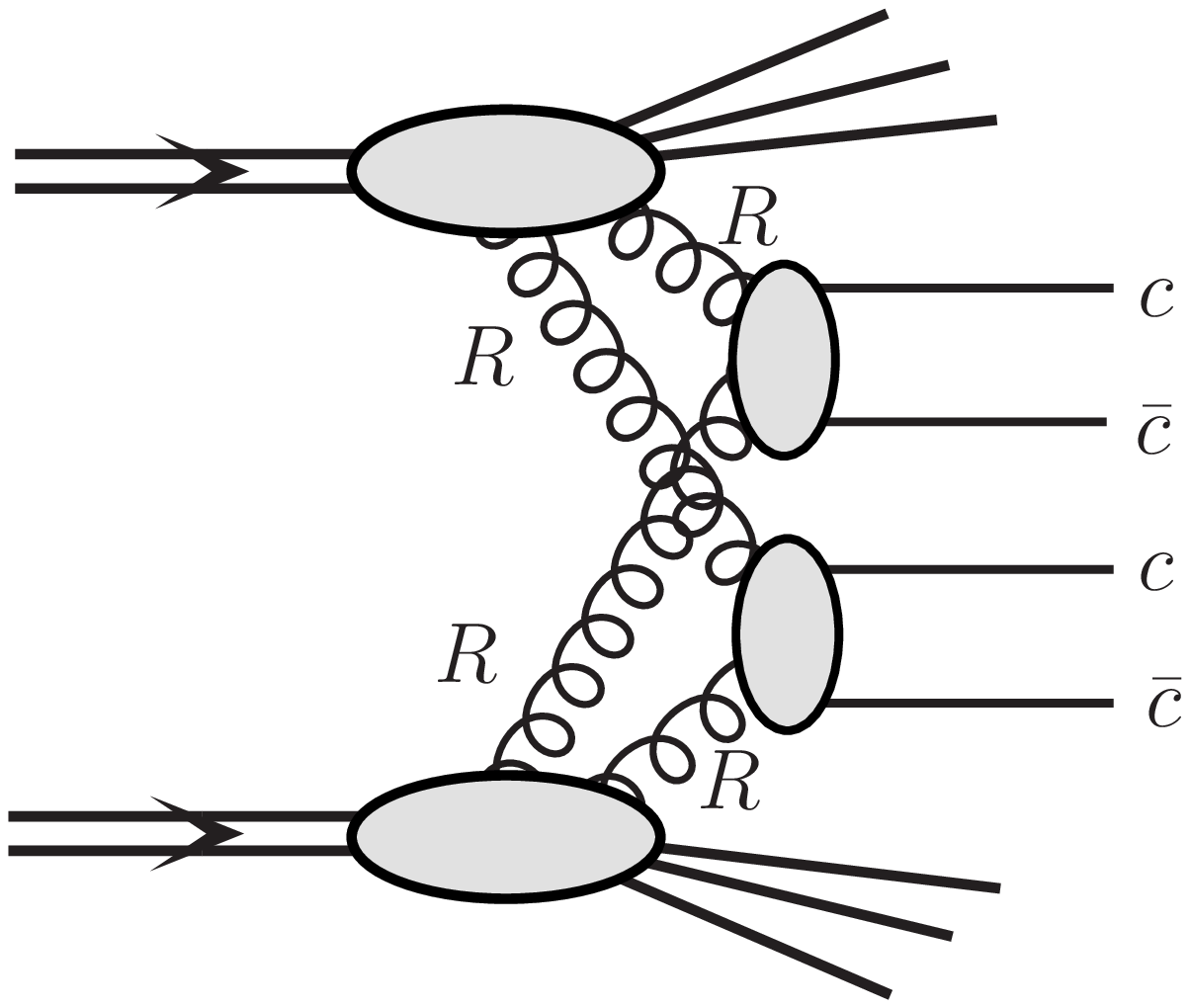}}
\end{minipage}
\hspace{0.0cm}
\begin{minipage}{0.24\textwidth}
 \centerline{\includegraphics[width=1.0\textwidth]{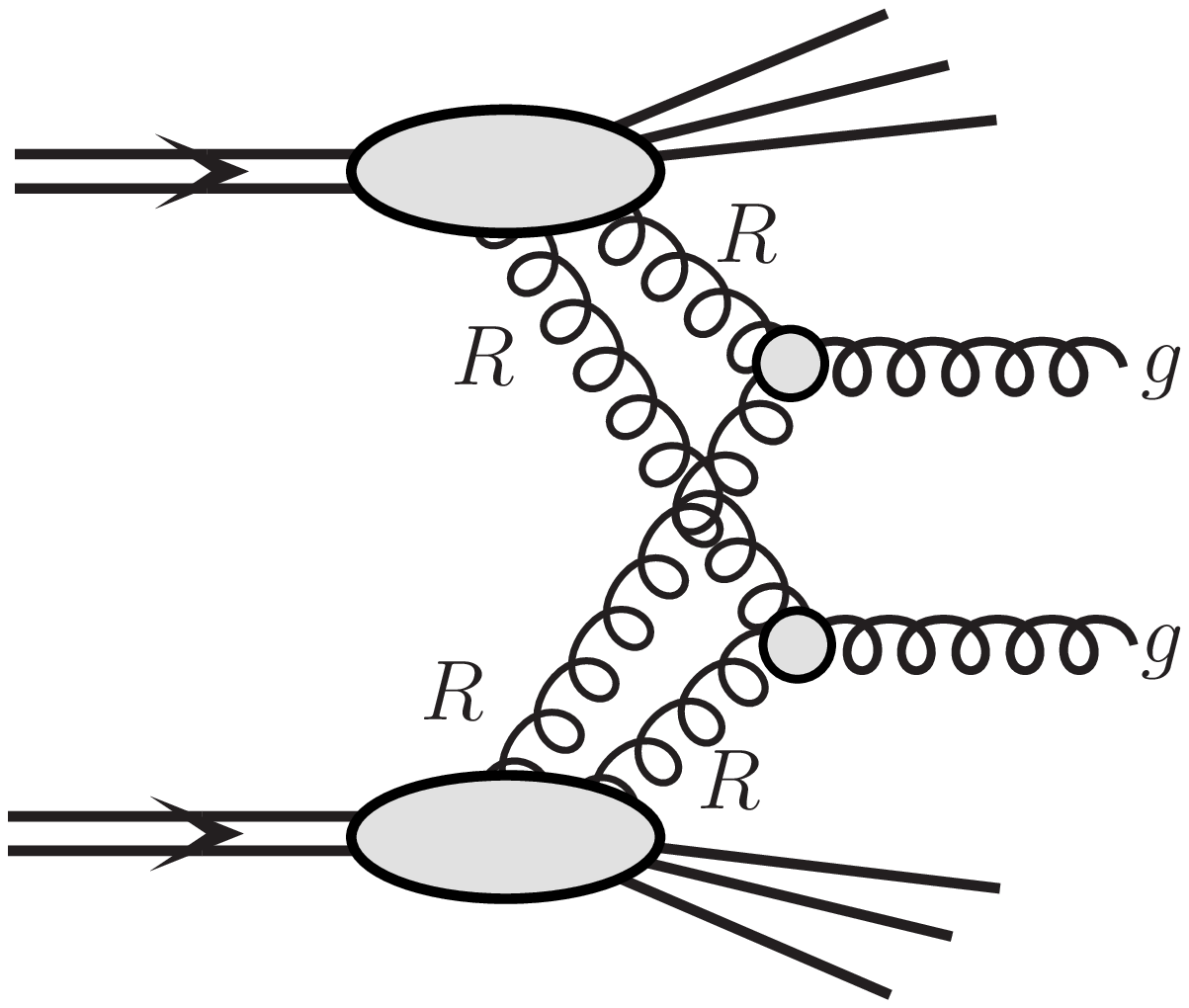}}
\end{minipage}
\vspace{0.0cm}
\begin{minipage}{0.24\textwidth}
 \centerline{\includegraphics[width=1.0\textwidth]{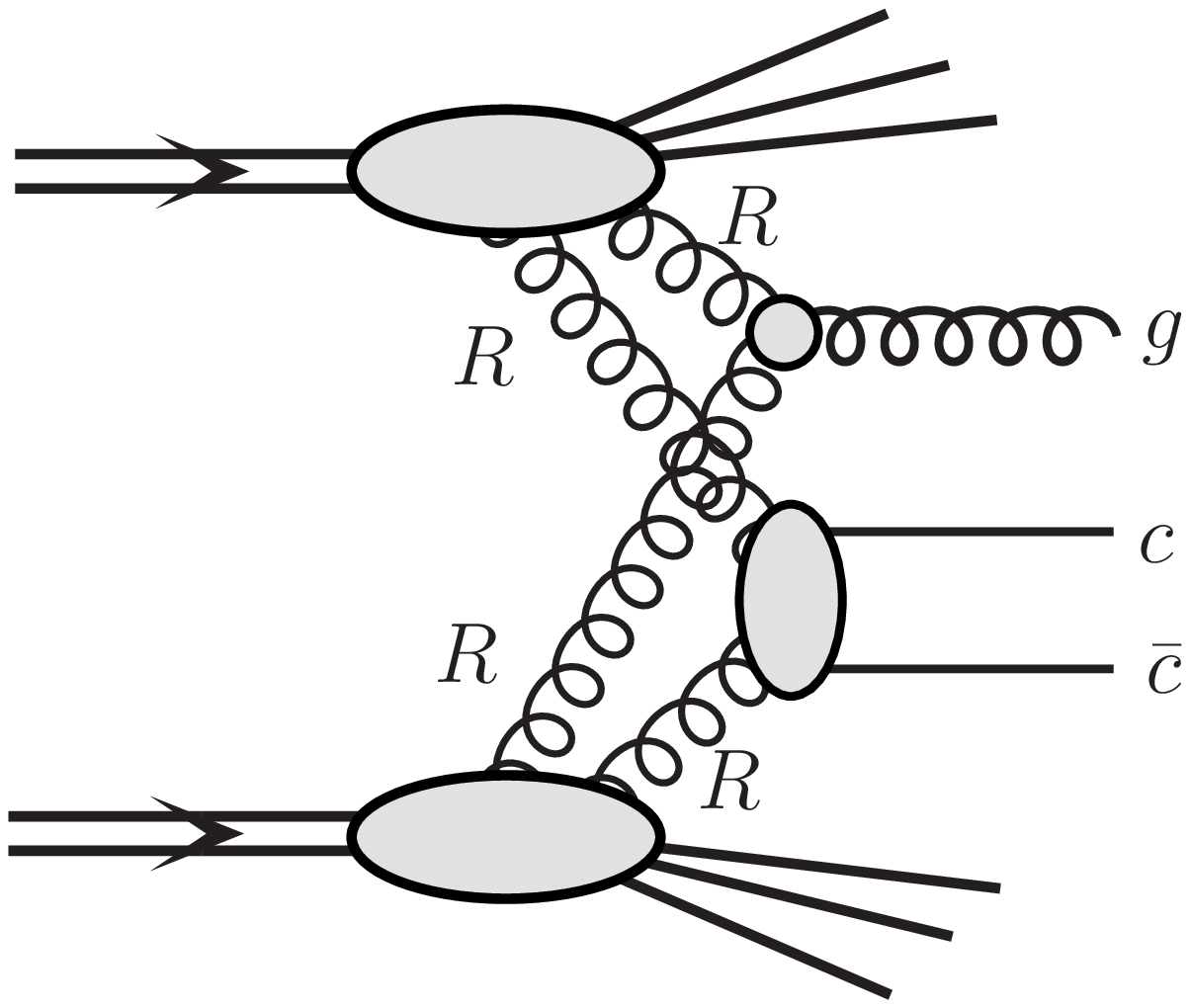}}
\end{minipage}
\hspace{0.0cm}
\begin{minipage}{0.24\textwidth}
 \centerline{\includegraphics[width=1.0\textwidth]{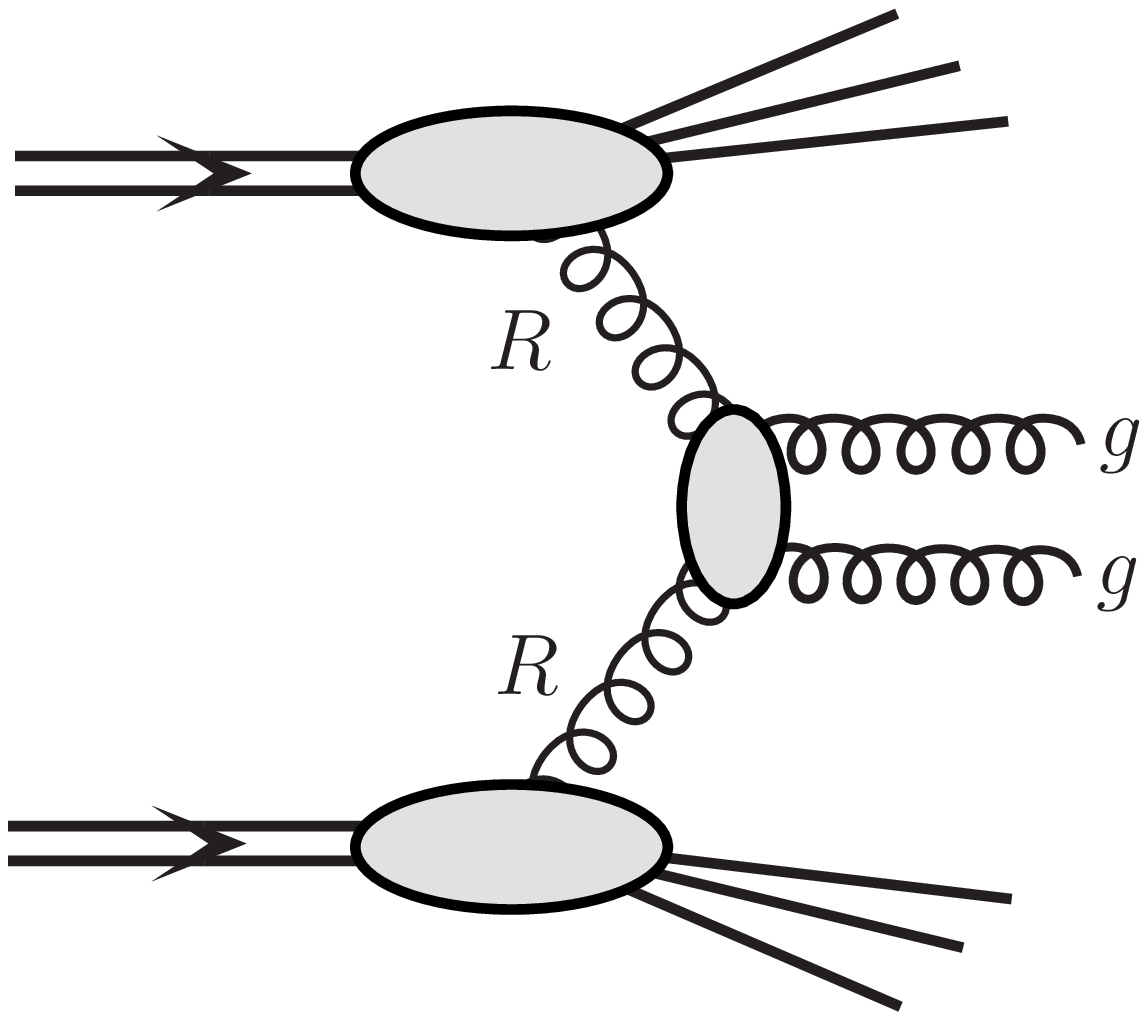}}
\end{minipage}
   \caption{
\small A diagrammatic illustration of the considered mechanisms.
 }
 \label{fig:diagrams}
 \end{center}
\end{figure}

We will compare numerical results for double $D$-meson production obtained with the two different fragmentation schemes.
In the (new) scenario with scale-dependent KKKS08 FFs and $g \to D$ fragmentation the number of contributing processes
grows compared to the standard (old) scenario with $c \to D$ fragmentation only. According to the new scenario one has to consider more processes for single $D$ meson production ($c$ and $g \to D$ components). This also causes an extension of the standard DPS $DD$ pair production by new mechanisms.
In addition to the coventional DPS $cc \to DD$ (left diagram in Fig.\ref{fig:diagrams})
considered in Refs.~\cite{Maciula:2013kd,vanHameren:2014ava,vanHameren:2015wva}
there is a double $g \to D$ (or double $g \to \bar D$) fragmentation mechanism, called here
DPS $gg \to DD$ (middle-left diagram in Fig.\ref{fig:diagrams}) as well as
the mixed DPS $gc \to DD$ contribution (middle-right diagram in Fig.\ref{fig:diagrams}).

As a consequence of the new approach a new SPS $gg \to DD$ mechanism shows up (right diagram in Fig.\ref{fig:diagrams}). In this case the two produced gluons are correlated in azimuth and the mechanism will naturally lead to an azimuthal correlation between the two $D$ mesons.
Such a correlation was actually observed in the LHCb experimental data \cite{Aaij:2012dz} and could not be explained by the SPS 2 $\to$ 4 perturbative $gg \to c \bar c c \bar c$ contribution (see e.g. Ref.~\cite{vanHameren:2015wva}) which is very small.

DPS cross section for production of $cc$, $gg$ or $gc$ system, assuming factorization of the
DPS model, can be written as:
\begin{eqnarray}
\frac{d \sigma^{DPS}(p p \to c  c  X)}{d y_1 d y_2 d^2
p_{1,t} d^2 p_{2,t}} = 
\frac{1}{2 \sigma_{eff}} \cdot \frac{d \sigma^{SPS}(p p \to c \bar c
X_1)}{d y_1  d^2 p_{1,t}} \cdot \frac{d \sigma^{SPS}(p p \to c \bar
c X_2)}{d y_2 d^2 p_{2,t}}, \label{DPScc_factorization_formula}
\end{eqnarray}
\begin{eqnarray}
\frac{d \sigma^{DPS}(p p \to g g  X)}{d y_1 d y_2 d^2
p_{1,t} d^2 p_{2,t}} =
\frac{1}{2 \sigma_{eff}} \cdot \frac{d \sigma^{SPS}(p p \to g
X_1)}{d y_1  d^2 p_{1,t}} \cdot \frac{d \sigma^{SPS}(p p \to g
X_2)}{d y_2 d^2 p_{2,t}}. \label{DPSgg_factorization_formula}
\end{eqnarray}
\begin{eqnarray}
\frac{d \sigma^{DPS}(p p \to g c  X)}{d y_1 d y_2 d^2
p_{1,t} d^2 p_{2,t}} =
\frac{1}{\sigma_{eff}} \cdot \frac{d \sigma^{SPS}(p p \to g
X_1)}{d y_1  d^2 p_{1,t}} \cdot \frac{d \sigma^{SPS}(p p \to c \bar
cX_2)}{d y_2 d^2 p_{2,t}}. \label{DPSgg_factorization_formula}
\end{eqnarray}

The often called pocket-formula is a priori a severe approximation. The flavour, spin and color correlations lead, in principle, to interference effects that result in its violation as discussed e.g. in Ref.~\cite{Diehl:2011yj}. Even for unpolarized proton beams, the spin polarization of the two partons from one hadron
can be mutually correlated, especially when the partons are relatively close in phase space (having comparable $x$'s). Moreover, in contrast to the standard single PDFs, the two-parton distributions have a nontrivial color structure which also may lead to a non-negligible correlations effects. 
Such effects are usually not included in phenomenological analyses. They were exceptionally discussed in the context of double charm production \cite{Echevarria:2015ufa}.
However, the effect on e.g. azimuthal correlations between charmed quarks was found there to be very small, much smaller than effects of the SPS
contribution associated with double gluon fragmentation discussed here.
In addition, including perturbative parton splitting mechanism also leads to a breaking of the pocket-formula \cite{Gaunt:2014rua}.
This formalism was so far formulated for the collinear leading-order
approach which for charm (double charm) may be a bit academic as this
leads to underestimation of the cross section.
Imposing sum rules also leads to a breaking of the factorized Ansatz
but the effect almost vanishes for small longitudinal momentum fractions 
\cite{Golec-Biernat:2015aza}. Taken the above we will use the pocket-formula in the
following.

All the considered mechanisms (see Fig.~\ref{fig:diagrams}) are calculated in the $k_t$-factorization approach with off-shell initial state partons and unintegrated ($k_{t}$-dependent) PDFs (unPDFs). Fully gauge invariant treatment of the initial-state off-shell
gluons and quarks can be achieved in the $k_t$-factorization approach only when
they are considered as Reggeized gluons or Reggeons. The relevant
Reggeized amplitudes can be presented using Fadin-Kuraev-Lipatov
effective vertices. The useful analytical formulae for $\overline{|{\cal M}_{RR \to g}|^2}$, $\overline{|{\cal M}_{R R
\rightarrow g g }|^2}$ and $\overline{|{\cal M}_{R R \rightarrow c
\bar c }|^2}$ squared amplitudes used in the calculations here can be found in Refs.~\cite{Nefedov:2014qea,Nefedov:2013ywa}.
We use the LO Kimber-Martin-Ryskin (KMR) unPDFs,
generated from the LO set of a up-to-date MMHT2014 collinear PDFs fitted also to the LHC data. In the perturbative part of the calculations we use a running LO $\alpha_{S}$
provided with the MMHT2014 PDFs. The charm quark mass is set to $m_{c} = 1.5$ GeV and the renormalization and factorization scales are equal to $\mu^{2} = p_{t}^{2}$ for $RR \to g$ subprocess, to the averaged transverse momentum $\mu^{2} = (p_{1t}^{2}+p_{2t}^{2})/2$ for $RR \to g g$, and to the averaged transverse mass $\mu^{2} = (m_{1t}^{2}+m_{2t}^{2})/2$ for $RR \to c \bar c$ case, where $m_{t} = \sqrt{p_{t}^{2} + m_{c}^{2}}$ (for more details see Ref.\cite{Maciula:2016wci}).

In order to calculate correlation observables for two mesons we
follow here, similar as in the single meson case, the fragmentation
function technique for hadronization process:
\begin{eqnarray}
\frac{d \sigma^{DPS}(pp \to D D X)}{d y_1 d y_{2} d^2 p_{1t}^{D} d^2 p_{2t}^{D}}
 &=&
\int \frac{D_{c \to D}(z_{1},\mu)}{z_{1}}\cdot \frac{D_{c \to D}(z_{2},\mu)}{z_{2}}\cdot
\frac{d \sigma^{DPS}(pp \to c c X)}{d y_1 d y_{2} d^2
  p_{1t}^{c} d^2 p_{2t}^{c}} d z_{1} d z_{2} \nonumber \\
  &+& \int \frac{D_{g \to D}(z_{1},\mu)}{z_{1}}\cdot \frac{D_{g \to D}(z_{2},\mu)}{z_{2}}\cdot
\frac{d \sigma^{DPS}(pp \to g g X)}{d y_1 d y_{2} d^2
  p_{1t}^{g} d^2 p_{2t}^{g}} d z_{1} d z_{2}\nonumber \\
  &+& \int \frac{D_{g \to D}(z_{1},\mu)}{z_{1}}\cdot \frac{D_{c \to D}(z_{2},\mu)}{z_{2}}\cdot
\frac{d \sigma^{DPS}(pp \to g c X)}{d y_1 d y_{2} d^2
  p_{1t}^{g} d^2 p_{2t}^{c}} d z_{1} d z_{2}, \nonumber \\
\end{eqnarray}
where: $p_{1t}^{g,c} = \frac{p_{1,t}^{D}}{z_{1}}$, $p_{2,t}^{g,c} =
  \frac{p_{2t}^{D}}{z_{2}}$ and meson momentum fractions $z_{1}, z_{2}\in (0,1)$.

The same formula for SPS $DD$-production via fragmentation of each of the gluon
reads
\begin{equation}
\frac{d \sigma^{SPS}_{gg}(pp \to D D X)}{d y_1 d y_{2} d^2 p_{1t}^{D} d^2 p_{2t}^{D}}
 \approx
\int \frac{D_{g \to D}(z_{1},\mu)}{z_{1}}\cdot \frac{D_{g \to D}(z_{2},\mu)}{z_{2}}\cdot
\frac{d \sigma^{SPS}(pp \to g g X)}{d y_1 d y_{2} d^2
  p_{1t}^{g} d^2 p_{2t}^{g}} d z_{1} d z_{2} \; ,
\end{equation}
where:
$p_{1t}^{g} = \frac{p_{1,t}^{D}}{z_{1}}$, $p_{2,t}^{g} =
  \frac{p_{2t}^{D}}{z_{2}}$ and
meson momentum fractions  $z_{1}, z_{2}\in (0,1)$.

\section{Comparison to the LHCb data}

Before we start a comparison of the theoretical results with the LHCb double charm data we wish to stress that
the both fragmentation schemes considered here lead to a very good (and very similar) description of the LHCb data for inclusive single $D$ meson production \cite{Maciula:2016wci}. So both prescriptions together with the $k_{t}$-factorization approach seem to be a good and legitimate starting points
for double charm production studies.  

\begin{figure}[!h]
\begin{center}
\begin{minipage}{0.4\textwidth}
 \centerline{\includegraphics[width=1.0\textwidth]{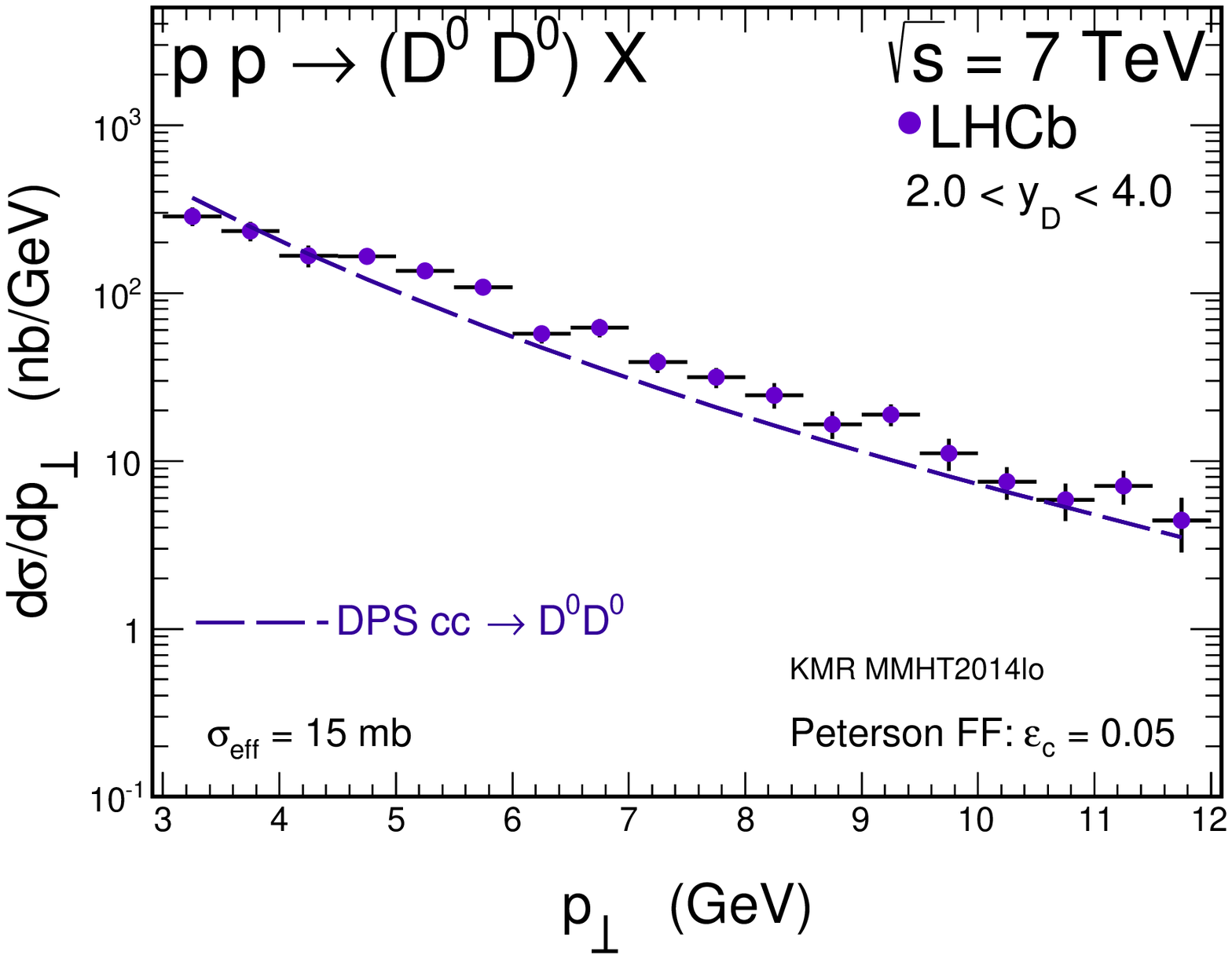}}
\end{minipage}
\hspace{0.5cm}
\begin{minipage}{0.4\textwidth}
 \centerline{\includegraphics[width=1.0\textwidth]{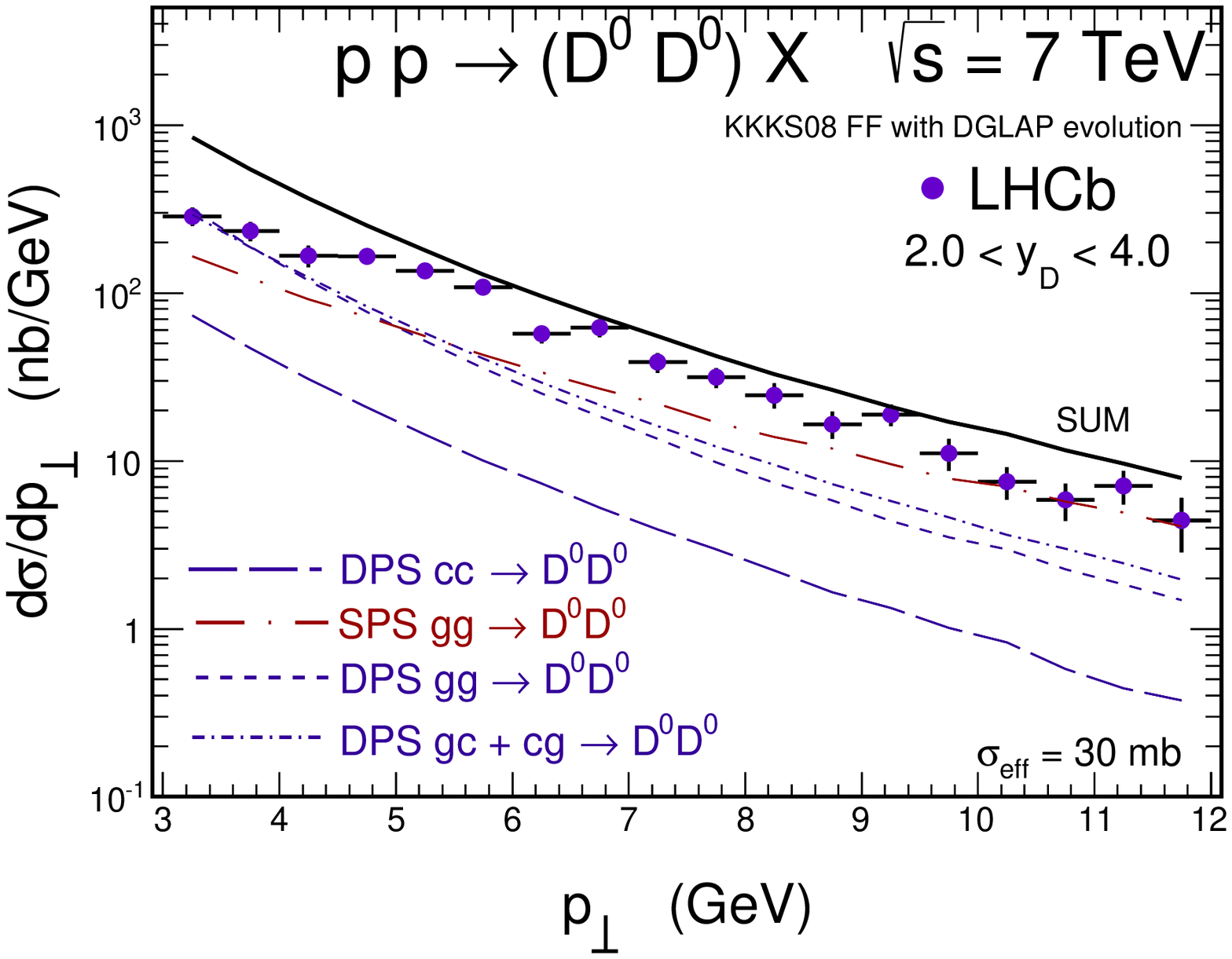}}
\end{minipage}
   \caption{
\small $D^0$ meson transverse momentum distribution within the LHCb
acceptance region.
The left panel is for the first scenario and for Peterson $c \to D$ fragmentation function
while the right panel is for the second scenario and for the fragmentation function
that undergo DGLAP evolution equation.
 }
 \label{fig:pT}
 \end{center}
\end{figure}
    
Now we wish to compare results of our theoretical approach for double charm production described
briefly in the previous section with the LHCb experimental data for $D^{0}D^{0}$ pair production.
In Fig.~\ref{fig:pT} we compare results of our calculation
with experimental distribution in transverse momentum of one of the
meson from the $D^{0}D^{0}$ pair. We show results for the first scenario when standard Peterson FF is used for the $c \to D^0$ fragmentation (left panel) as well as
the result for the second scenario when the KKKS08 FFs with DGLAP evolution for $c \to D^0$ and $g \to D^0$ are used. 
One can observe that the DPS $cc \to D^{0}D^{0}$ contribution in the new scenario is
much smaller than in the old scenario. In addition, the slope of the
distribution in transverse momentum changes. Both the effects are due to evolution of
corresponding FF in the second scenario, compared to
lack of such an effect in the first scenario.
The different new mechanisms shown in Fig.~\ref{fig:diagrams}
give contributions of similar size. We can obtain a better agreement in the second case provided $\sigma_{eff}$ parameter is increased from $15$ mb to $30$ mb. Even then we overestimate the LHCb data for $3 < p_{T} < 5$ GeV.

\begin{figure}[!h]
\begin{center}
\begin{minipage}{0.4\textwidth}
 \centerline{\includegraphics[width=1.0\textwidth]{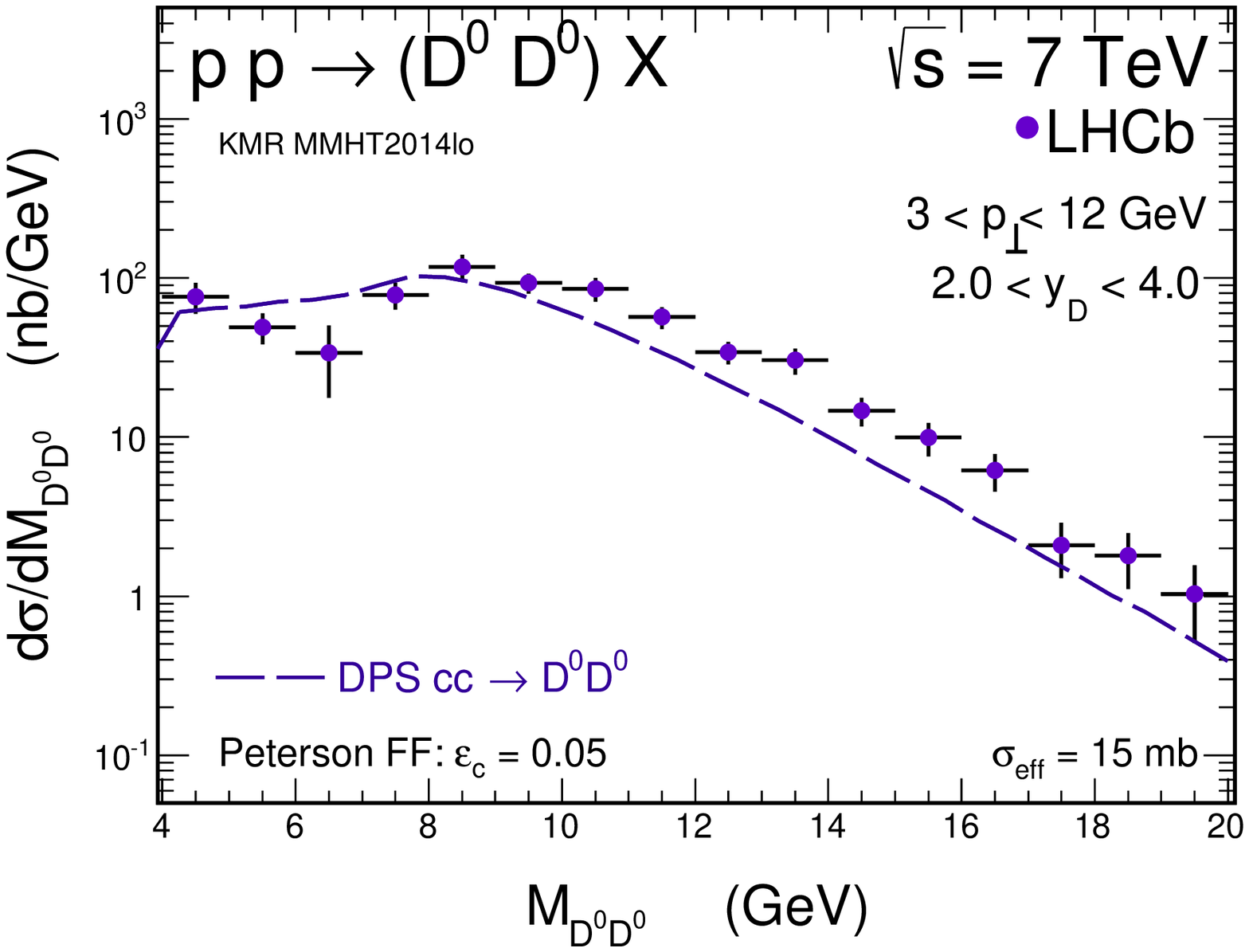}}
\end{minipage}
\hspace{0.5cm}
\begin{minipage}{0.4\textwidth}
 \centerline{\includegraphics[width=1.0\textwidth]{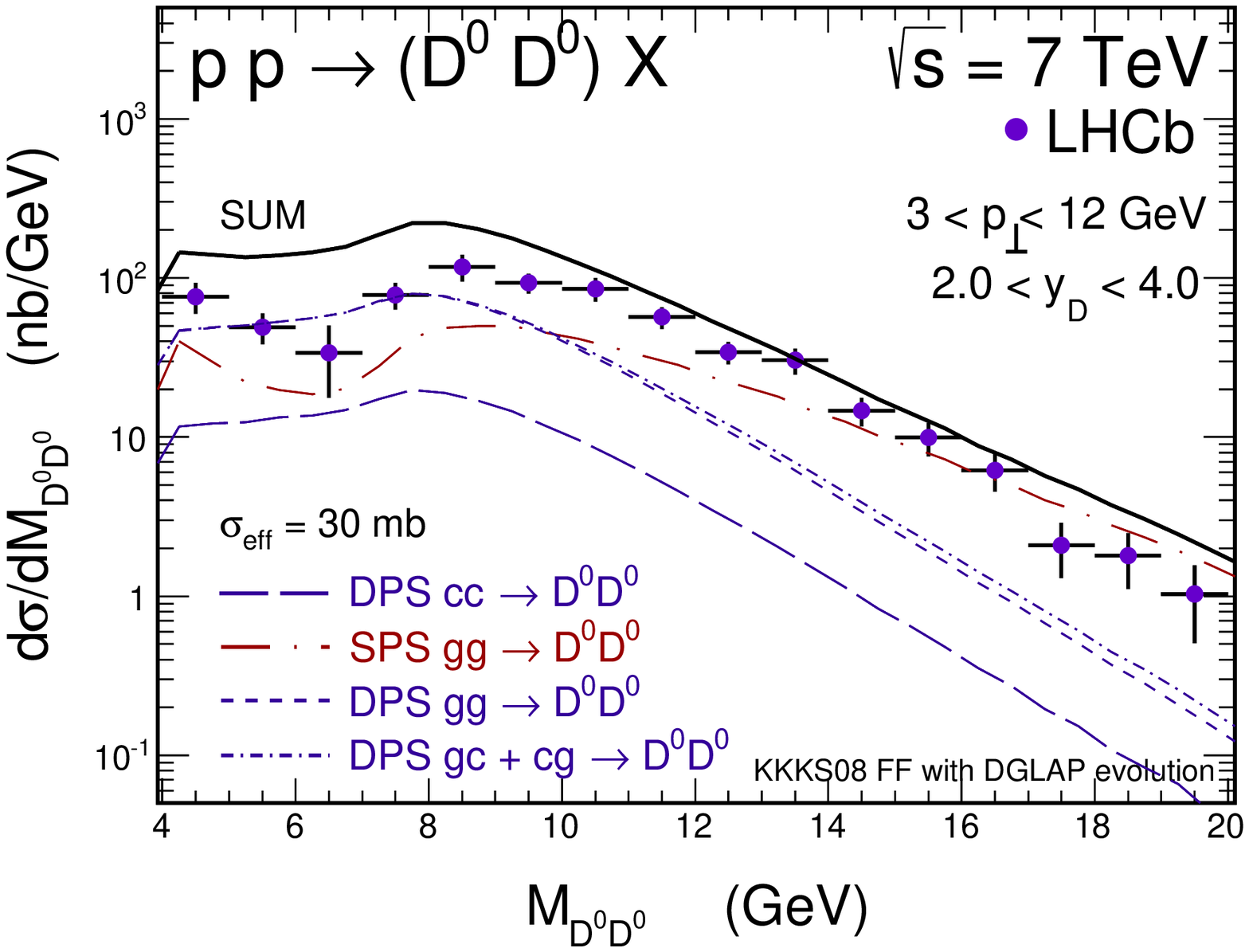}}
\end{minipage}
   \caption{
\small The same as in the previous figure but for $M_{D^0 D^0}$ dimeson invariant mass distribution.
 }
 \label{fig:Minv}
\end{center}
\end{figure}
\begin{figure}[!h]
\begin{center}
\begin{minipage}{0.4\textwidth}
 \centerline{\includegraphics[width=1.0\textwidth]{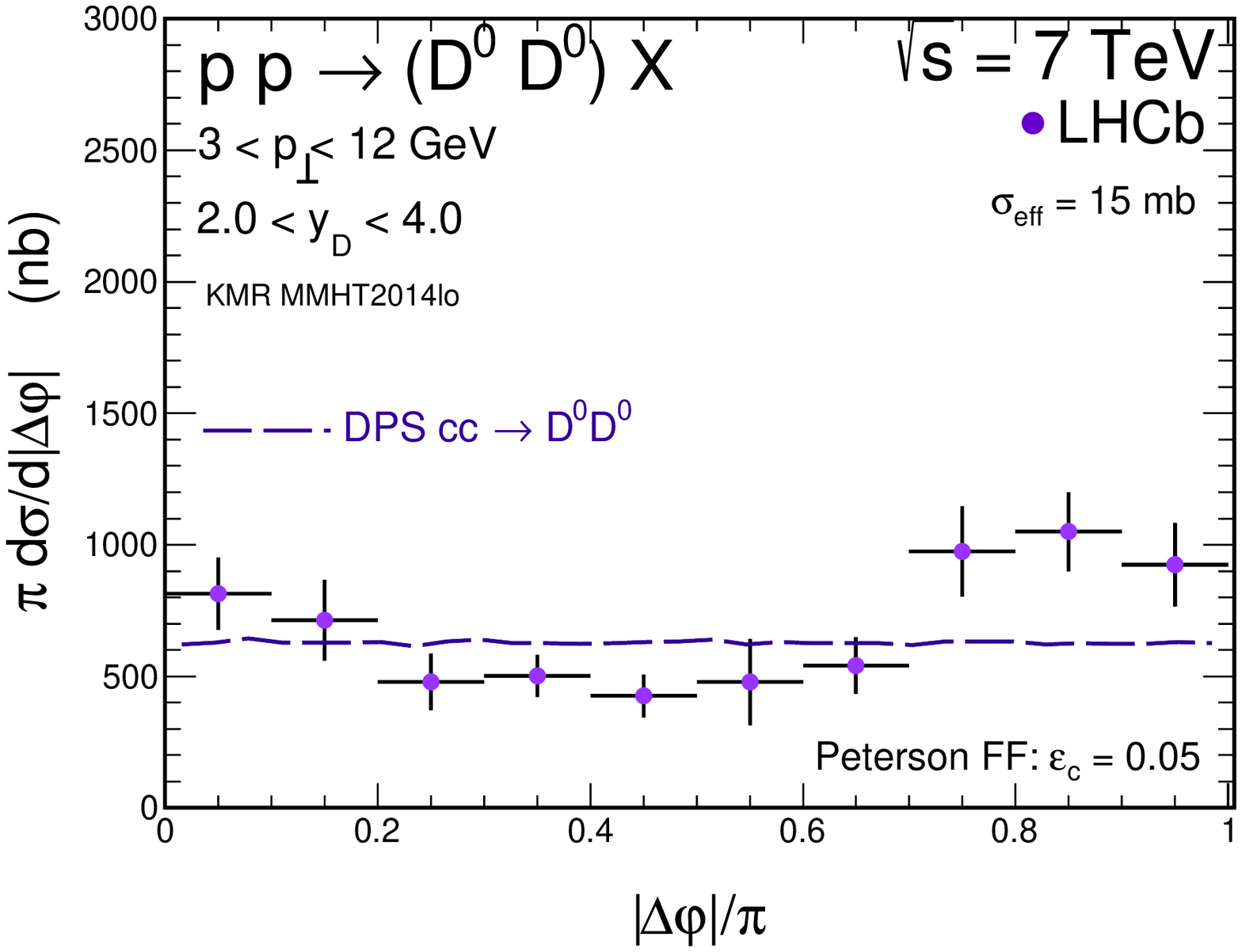}}
\end{minipage}
\hspace{0.5cm}
\begin{minipage}{0.4\textwidth}
 \centerline{\includegraphics[width=1.0\textwidth]{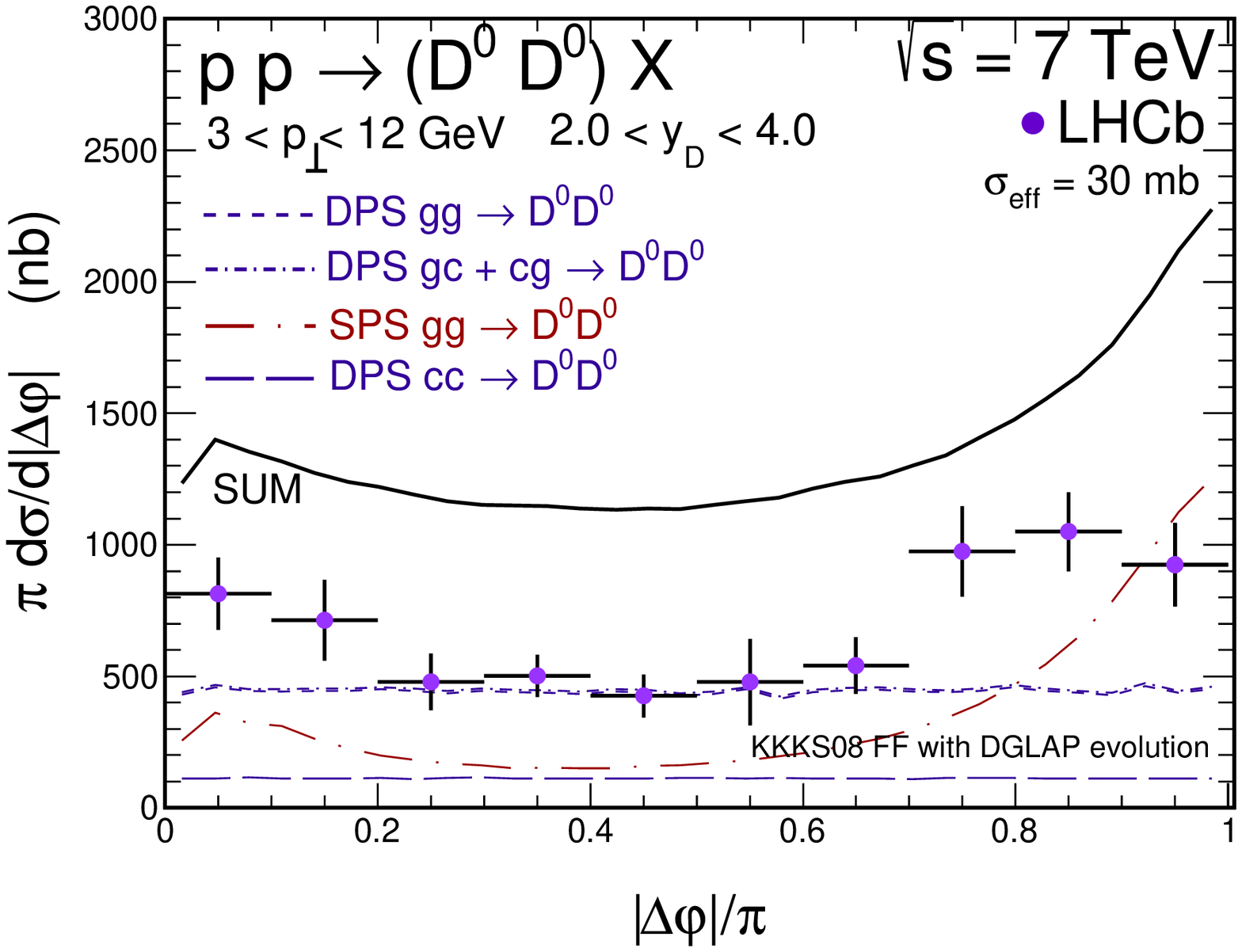}}
\end{minipage}
   \caption{
\small The same as in the previous figure but for the distribution in azimuthal angle $\varphi_{D^0 D^0}$.
 }
 \label{fig:phid}
\end{center}
\end{figure}

In Fig.~\ref{fig:Minv} we show dimeson invariant mass distribution $M_{D^0 D^0}$
again for the two cases considered. In the first scenario we get a good agreement only for small invariant 
masses while in the second scenario we get a good agreement
only for large invariant masses. The large invariant masses are
strongly correlated with large transverse momenta, so the situation 
here (for the invariant mass distribution) is quite similar as 
in Fig.~\ref{fig:pT} for the transverse momentum distribution.

In Fig.~\ref{fig:phid} we show azimuthal angle correlation $\varphi_{D^0 D^0}$ 
between $D^0$ and $D^0$.
While the correlation function in the first scenario is completely flat,
the correlation function in the second scenario shows some tendency similar as in the experimental data.
The observed overestimation comes from the region of small transverse momenta.

\section{Conclusions}

In the present paper we have discussed production of $D^0D^0$ pairs in proton-proton collisions at the LHC. We have considered
the DPS mechanism of double $c \bar c$ production
and subsequent double hadronization of two $c$ quarks or two $\bar c$
antiquarks using $c \to D^0$ or $c \to {\bar D}^0$ FFs that undergo DGLAP evolution.
Furthermore, we have included also production of $gg$ (both SPS and DPS) and DPS $gc$ final states and their
subsequent hadronization to the neutral pseudoscalar $D$ mesons.

When added together the new mechanisms with adjusted $\sigma_{eff}$ give similar result as in the first scenario
with one subprocess ($cc \to DD$) and scale-independent FF.
However, some correlation observables, such as dimeson invariant mass
or azimuthal correlations between $D$ mesons, are slightly better
described.

In our calculation, within the second scenario a larger value 
of $\sigma_{eff}$ is needed to describe the LHCb data than found from the review of several 
experimental studies of different processes.
This can be partially understood by a lower contribution of
perturbative-parton splitting as found in Ref.~\cite{Gaunt:2014rua}
and/or due to nonperturbative correlations in the nucleon which may lead
to transverse momentum dependent $\sigma_{eff}$.
Clearly more involved studies are needed to understand the situation
in detail. Some problem may be also related to the fact
that the FFs used in the second scenario
were obtained in the DGLAP formalism with massless $c$ quarks and 
$\bar c$ antiquarks which may be a too severe approximation,
especially for low factorization scales (i.e. low transverse momenta).
We expect that including mass effects in the evolution would lower
the $g \to c$ fragmentation.

The presence of the new SPS mechanism may mean that the extraction
of $\sigma_{eff}$ directly from the LHCb experimental data \cite{Aaij:2012dz}
may be not correct.

For more references and details of the calculations presented here we refer the reader to our regular article \cite{Maciula:2016wci}.

\end{document}